\documentclass[aps,prl,twocolumn,superscriptaddress,showpacs,preprintnumbers,amsmath,amssymb]{revtex4}

\usepackage{graphicx,color}
\usepackage{dcolumn}
\usepackage{subfigure}
\usepackage{multirow}

\graphicspath{{ps}}

\begin{document}

\preprint{\vbox{ \hbox{   }
			\hbox{Belle Preprint {\it 2011-04}}
			\hbox{KEK Preprint {\it 2010-52}}
}}

\title{ \quad\\[1.0cm]\boldmath Evidence for the Suppressed Decay 
$B^- \rightarrow DK^-$, $D \rightarrow K^+\pi^-$}

\affiliation{Budker Institute of Nuclear Physics, Novosibirsk}
\affiliation{Faculty of Mathematics and Physics, Charles University, Prague}
\affiliation{University of Cincinnati, Cincinnati, Ohio 45221}
\affiliation{Department of Physics, Fu Jen Catholic University, Taipei}
\affiliation{Gifu University, Gifu}
\affiliation{Hanyang University, Seoul}
\affiliation{University of Hawaii, Honolulu, Hawaii 96822}
\affiliation{High Energy Accelerator Research Organization (KEK), Tsukuba}
\affiliation{University of Illinois at Urbana-Champaign, Urbana, Illinois 61801}
\affiliation{Indian Institute of Technology Guwahati, Guwahati}
\affiliation{Institute of High Energy Physics, Chinese Academy of Sciences, Beijing}
\affiliation{Institute of High Energy Physics, Vienna}
\affiliation{Institute of High Energy Physics, Protvino}
\affiliation{Institute for Theoretical and Experimental Physics, Moscow}
\affiliation{J. Stefan Institute, Ljubljana}
\affiliation{Kanagawa University, Yokohama}
\affiliation{Institut f\"ur Experimentelle Kernphysik, Karlsruher Institut f\"ur Technologie, Karlsruhe}
\affiliation{Korea Institute of Science and Technology Information, Daejeon}
\affiliation{Korea University, Seoul}
\affiliation{Kyungpook National University, Taegu}
\affiliation{\'Ecole Polytechnique F\'ed\'erale de Lausanne (EPFL), Lausanne}
\affiliation{Faculty of Mathematics and Physics, University of Ljubljana, Ljubljana}
\affiliation{University of Maribor, Maribor}
\affiliation{Max-Planck-Institut f\"ur Physik, M\"unchen}
\affiliation{University of Melbourne, School of Physics, Victoria 3010}
\affiliation{Nagoya University, Nagoya}
\affiliation{Nara Women's University, Nara}
\affiliation{National Central University, Chung-li}
\affiliation{National United University, Miao Li}
\affiliation{Department of Physics, National Taiwan University, Taipei}
\affiliation{H. Niewodniczanski Institute of Nuclear Physics, Krakow}
\affiliation{Nippon Dental University, Niigata}
\affiliation{Niigata University, Niigata}
\affiliation{University of Nova Gorica, Nova Gorica}
\affiliation{Novosibirsk State University, Novosibirsk}
\affiliation{Osaka City University, Osaka}
\affiliation{Panjab University, Chandigarh}
\affiliation{Research Center for Nuclear Physics, Osaka University, Osaka}
\affiliation{Saga University, Saga}
\affiliation{University of Science and Technology of China, Hefei}
\affiliation{Seoul National University, Seoul}
\affiliation{Sungkyunkwan University, Suwon}
\affiliation{School of Physics, University of Sydney, NSW 2006}
\affiliation{Tata Institute of Fundamental Research, Mumbai}
\affiliation{Excellence Cluster Universe, Technische Universit\"at M\"unchen, Garching}
\affiliation{Tohoku Gakuin University, Tagajo}
\affiliation{Tohoku University, Sendai}
\affiliation{Department of Physics, University of Tokyo, Tokyo}
\affiliation{Tokyo Institute of Technology, Tokyo}
\affiliation{Tokyo Metropolitan University, Tokyo}
\affiliation{Tokyo University of Agriculture and Technology, Tokyo}
\affiliation{CNP, Virginia Polytechnic Institute and State University, Blacksburg, Virginia 24061}
\affiliation{Yonsei University, Seoul}
  \author{Y.~Horii}\affiliation{Tohoku University, Sendai} 
  \author{K.~Trabelsi}\affiliation{High Energy Accelerator Research Organization (KEK), Tsukuba} 
  \author{H.~Yamamoto}\affiliation{Tohoku University, Sendai} 
  \author{I.~Adachi}\affiliation{High Energy Accelerator Research Organization (KEK), Tsukuba} 
  \author{H.~Aihara}\affiliation{Department of Physics, University of Tokyo, Tokyo} 
  \author{K.~Arinstein}\affiliation{Budker Institute of Nuclear Physics, Novosibirsk}\affiliation{Novosibirsk State University, Novosibirsk} 
  \author{V.~Aulchenko}\affiliation{Budker Institute of Nuclear Physics, Novosibirsk}\affiliation{Novosibirsk State University, Novosibirsk} 
  \author{T.~Aushev}\affiliation{\'Ecole Polytechnique F\'ed\'erale de Lausanne (EPFL), Lausanne}\affiliation{Institute for Theoretical and Experimental Physics, Moscow} 
  \author{V.~Balagura}\affiliation{Institute for Theoretical and Experimental Physics, Moscow} 
  \author{E.~Barberio}\affiliation{University of Melbourne, School of Physics, Victoria 3010} 
% ...
\collaboration{Belle Collaboration}

\begin{abstract}
The suppressed decay chain $B^- \rightarrow DK^-$, $D\rightarrow K^+\pi^-$,
where $D$ indicates a $\bar{D}^0$ or $D^0$ state,
provides important information on the $CP$-violating angle $\phi_3$.
We measure the ratio ${\cal R}_{DK}$ of the decay rates
to the favored mode $B^- \rightarrow DK^-$, $D\rightarrow K^-\pi^+$
to be ${\cal R}_{DK} = [1.63^{+0.44}_{-0.41}({\rm stat})^{+0.07}_{-0.13}({\rm syst})] \times10^{-2}$,
which indicates the first evidence of the signal with a significance of $4.1\sigma$.
We also measure the asymmetry ${\cal A}_{DK}$ between the charge-conjugate decays
to be ${\cal A}_{DK} = -0.39^{+0.26}_{-0.28}({\rm stat})^{+0.04}_{-0.03}({\rm syst})$.
The results are based on the full $772 \times 10^6~B\bar{B}$ pair data sample
collected at the $\Upsilon(4S)$ resonance with the Belle detector.
\end{abstract}


\maketitle

\tighten

{\renewcommand{\thefootnote}{\fnsymbol{footnote}}}
\setcounter{footnote}{0}

Determinations of the parameters of the standard model are fundamentally important;
any significant discrepancy between the expected and measured values
would be a signature of new physics.
The Cabibbo-Kobayashi-Maskawa matrix~\protect\cite{Cabibbo, KM} consists
of weak interaction parameters for the quark sector, one of which is
the $CP$-violating angle $\phi_3 \equiv \arg{(-{V_{ud}}{V_{ub}}^{*}/{V_{cd}}{V_{cb}}^{*})}$~\protect\cite{gamma}.
Several methods proposed for measuring $\phi_3$ exploit interference
in the decay $B^-\rightarrow DK^-$ ($D = \bar{D}^0~{\rm or}~D^0$),
where the two $D$ states decay to a common final state~\protect\cite{DK, GLW, ADS, Dalitz}.
One of the methods utilizes
the decay $B^- \rightarrow DK^-$, $D \rightarrow K^+\pi^-$~\protect\cite{ADS}.
The magnitudes of interfering amplitudes are comparable
and hence can enhance the effects of $CP$ violation.
Previous studies of this decay mode have 
not found a significant signal yield~\protect\cite{ADS_Belle, ADS_BaBar}.


In this analysis, we measure the ratio ${\cal R}_{DK}$
of the aforementioned suppressed decay to the favored decay,
$B^- \rightarrow DK^-$, $D \rightarrow K^-\pi^+$,
and the $CP$ asymmetry ${\cal A}_{DK}$ defined as
\begin{eqnarray}
{\cal R}_{DK} &\equiv& \frac{{\cal B}([K^+\pi^-]_DK^-)+{\cal B}([K^-\pi^+]_DK^+)}
{{\cal B}([K^-\pi^+]_DK^-)+{\cal B}([K^+\pi^-]_DK^+)}, \label{eq:rdk} \\
{\cal A}_{DK} &\equiv& \frac{{\cal B}([K^+\pi^-]_DK^-)-{\cal B}([K^-\pi^+]_DK^+)}
{{\cal B}([K^+\pi^-]_DK^-)+{\cal B}([K^-\pi^+]_DK^+)}, \label{eq:adk}
\end{eqnarray}
where $[f]_D$ indicates the final state $f$ originating from a $\bar{D}^0$ or $D^0$ meson.
The same selection criteria and fitting functions are used
for the suppressed decays and the favored decays wherever possible
in order to cancel systematic uncertainties.
The observables are related to $\phi_3$ as follows:
\begin{eqnarray}
  {\cal R}_{DK} &=& r_B^2 + r_D^2 + 2r_Br_D \cos{(\delta_B+\delta_D)}\cos{\phi_3}, \label{eq:rdk_rel} \\
  {\cal A}_{DK} &=& 2r_Br_D\sin{(\delta_B+\delta_D)}\sin{\phi_3}/{\cal R}_{DK}, \label{eq:adk_rel}
\end{eqnarray}
where $r_B = |A(B^-\rightarrow \bar{D}^0K^-)/A(B^-\rightarrow D^0K^-)|$,
$r_D = |A(D^0\rightarrow K^+\pi^-)/A(\bar{D}^0\rightarrow K^+\pi^-)|$,
and $\delta_B$ ($\delta_D$) is the strong phase difference between
the two $B$ ($D$) decay amplitudes appearing in the ratios.
By combining other experimental inputs~\protect\cite{GLW_Belle, GLW_BaBar, HFAG},
the value of $\phi_3$ can be extracted in a model-independent manner~\protect\cite{GLW, ADS}.
The decay $B^-\rightarrow D \pi^-$, $D\rightarrow K^+\pi^-$ is also analyzed as a reference.
For this final state the decay rate is relatively large whereas the effect of $\phi_3$ is small.

The results are based on the full $772 \times 10^6~B\bar{B}$ pair data sample
collected at the $\Upsilon(4S)$ resonance
with the Belle detector located at the KEKB asymmetric-energy $e^+e^-$
collider~\protect\cite{KEKB}.
The Belle detector is described in detail elsewhere~\protect\cite{Belle}.
The primary detector components used in this analysis are a tracking system
consisting of a silicon vertex detector and a 50-layer central drift chamber (CDC),
an array of aerogel threshold Cherenkov counters (ACC),
and a barrel-like arrangement of time-of-flight scintillation counters (TOF).

Neutral $D$ meson candidates are reconstructed from pairs of oppositely charged tracks.
For each track, we apply a particle identification requirement based
on information from the ACC and TOF,
and specific ionization measurements from the CDC.
The efficiency to identify a kaon or a pion is 85--95\%, 
while the probability of misidentifying a pion (kaon) as a kaon (pion) is 10--20\%.
The invariant mass of the $K \pi$ pair is required
to satisfy $1.850~{\rm GeV}/c^2 < M(K\pi) < 1.880~{\rm GeV}/c^2$,
which corresponds to approximately $\pm 3\sigma$ around the world-average value
of the $D$ mass~\protect\cite{PDG},
where $\sigma$ denotes the resolution in $M(K\pi)$.
To improve the momentum determination, tracks from the $D$ candidate are refitted with
their invariant mass constrained to the nominal $D$ mass.

$B$ meson candidates are reconstructed
by combining a $D$ candidate with a charged hadron candidate.
The signal is identified using the beam-energy-constrained mass ($M_{\rm bc}$)
and the energy difference ($\Delta E$) defined, in the $e^+e^-$ center-of-mass frame, as
$M_{\rm bc} = \sqrt{\mathstrut E_{\rm beam}^2 - |\vec{p}_B|^2}$
and $\Delta E = E_B - E_{\rm beam}$, 
where $E_{\rm beam}$ is the beam energy
and $\vec{p}_B$ and $E_B$ are the momentum and energy, respectively, of the $B$ meson candidate.
We require $5.271~{\rm GeV}/c^2 < M_{\rm bc} < 5.287~{\rm GeV}/c^2$,
which corresponds to $\pm 3\sigma$ around the $B$ mass value~\protect\cite{PDG}
with $\sigma$ denoting the resolution in $M_{\rm bc}$.

The dominant backgrounds arise from the continuum processes $e^+e^-\rightarrow q\bar{q}$ ($q = u,d,s,c$).
In order to remove $D^{*\pm}\rightarrow D\pi^\pm$ decays produced in such a process,
we employ the variable $\Delta M$ defined as the mass difference
between the $D^{*\pm}$ and $D$ candidates,
where the $D^{*\pm}$ candidate is reconstructed from the $D$ candidate used in the $B$ reconstruction
and a $\pi^\pm$ candidate not used in the $B$ reconstruction.
We require $\Delta M > 0.15 ~ {\rm GeV}/c^{2}$, which removes 28\% of the $c\bar{c}$ background
as well as some of the $B\bar{B}$ background.
The loss of signal efficiency is 1.4\%.

The $q\bar{q}$ background is further discriminated
with a neural network technique based on the NeuroBayes package~\protect\cite{NeuroBayes}.
The inputs to the network are:
1) a Fisher discriminant~\protect\cite{Fisher} formed from modified Fox-Wolfram moments~\protect\cite{KSFW},
2) the vertex separation between the $B$ candidate and the remaining tracks,
3) the cosine of the decay angle of $D\rightarrow K^+\pi^-$,
where the decay angle is defined as the angle between
the $K^+$ candidate and the $B^-$ candidate in the rest frame of the $D$,
4) the cosine of the angle between the $B$ candidate and the beam axis
in the $e^+e^-$ center-of-mass frame,
5) the expected $B$ flavor dilution factor
that ranges from zero for no flavor information
to unity for unambiguous flavor assignment~\protect\cite{flavor_tag},
6) the cosine of the angle between the thrust axis
of the $B$ candidate and that of the rest of the event,
where the thrust axis is oriented in such a way that the sum of momentum projections is maximized,
and four other variables that exploit the kinematics of the events.
The neural network is trained with Monte Carlo (MC) events.
A requirement is applied on the network output ($\it NB$)
that preserves 96\% of the signal while rejecting 74\% of the background.

There are a few background modes that can peak in the signal window (`peaking background').
The decay $B^- \rightarrow [K^+K^-]_D\pi^-$ may contribute to the background
for $B^- \rightarrow [K^+\pi^-]_DK^-$ if the $D$ candidate is misreconstructed.
To reject this background,
we veto candidates satisfying $1.840~{\rm GeV}/c^2 < M(KK) <1.890~{\rm GeV}/c^2$.
The favored decay $B^- \rightarrow [K^-\pi^+]_Dh^-$ ($h=K$ or $\pi$) can also produce a 
peaking background for the suppressed decay if both the kaon and 
the pion from the $D$ decay are misidentified and the particle assignments are interchanged.
We thus veto candidates for which the invariant mass of the $K\pi$ pair is inside
a $(1.865 \pm 0.020)~{\rm GeV}/c^2$ window when the mass assignments are exchanged.

The signal yield is extracted from the two-dimensional distribution of $\Delta E$ and $\it NB$
using an extended unbinned maximum likelihood fit.
The fit is simultaneously applied to $DK^-$, $DK^+$, $D\pi^-$ and $D\pi^+$.
The components of the fit are divided into signal, `feed-across',
$B\bar{B}$ background, and $q\bar{q}$ background,
as described in detail below.
For each component,
the correlation between $\Delta E$ and $\it NB$ is found to be small.
We thus obtain the probability density function (PDF)
by taking a product of individual PDFs for $\Delta E$ and $\it NB$.
For $\it NB$, we use one-dimensional histogram PDFs for all components.

For the $\Delta E$ signal PDF, we use a sum of two Gaussians
whose parameters are fixed from the data for the favored modes.
For $\it NB$, we obtain the PDF by applying $|\Delta E|<0.01~{\rm GeV}$ to the same samples.

The $D\pi$ ($DK$) feed-across is the background from misidentified $D\pi$ ($DK$) final states
that peaks in the fit to the $DK$ ($D\pi$) sample.
The shift due to the incorrect mass assignment makes
the $\Delta E$ distribution asymmetric,
and thus we use a sum of two asymmetric Gaussians,
for which the left and right sides have different widths.
The corresponding parameters for the $D\pi$ feed-across
are obtained from the favored mode in data,
while those for the $DK$ feed-across
are obtained from MC sample because of low statistics in the data.
The PDFs for $\it NB$
are obtained from the same reference samples used for $\Delta E$ calibration
after applying the additional requirement $|\Delta E-0.05~{\rm GeV}|<0.01~{\rm GeV}$ to the data sample.
The $K/\pi$ misidentification probabilities are fixed from the samples of the favored modes.

The $B\bar{B}$ background populates the entire $\Delta E$ region.
This background is fitted with an exponential PDF that models the tails of the backgrounds from the modes
$B^- \rightarrow D^*\pi^-$, $B^- \rightarrow D\rho^-$, and $B^-\rightarrow D^*K^-$,
which peak in the negative $\Delta E$ region,
as well as combinatorial backgrounds.
The $\it NB$ PDF is obtained from $B\bar{B}$ MC samples, 
in which all known $B$ and $\bar{B}$ meson decays are included.
The charge asymmetry for the $B\bar{B}$ background is floated in the fit.

The $q\bar{q}$ background distribution in $\Delta E$ is modeled by a linear function.
The PDF of $\it NB$ is obtained from a sideband sample of data:
$5.20~{\rm GeV}/c^2 < M_{\rm bc} < 5.26~{\rm GeV}/c^2$
and $0.15~{\rm GeV} < \Delta E < 0.30~{\rm GeV}$.
The charge asymmetry for the $q\bar{q}$ background is fixed to zero.

\begin{figure}[htb]
 \begin{center}
  \includegraphics[width=8.4cm,height=8.4cm]{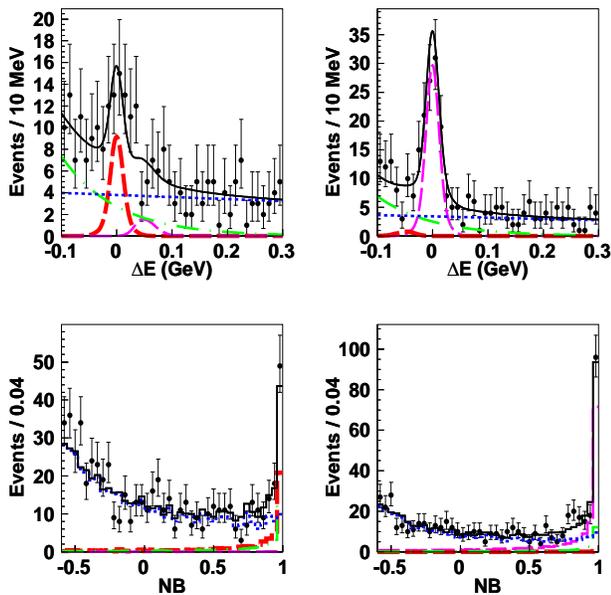}
  \caption{$\Delta E$ (${\it NB}>0.9$) and $\it NB$ ($|\Delta E|<0.03~{\rm GeV}$) distributions
for $[K^+\pi^-]_DK^-$ (left) and $[K^+\pi^-]_D\pi^-$ (right).
Charge-conjugate decays are included. In these plots,
$[K^+\pi^-]_DK^-$ components are shown by thicker dashed curves (red),
and $[K^+\pi^-]_D\pi^-$ components are shown by thinner dashed curves (magenta).
$B\bar{B}$ backgrounds are shown by dash-dotted curves (green)
while $q\bar{q}$ backgrounds are shown by dotted curves (blue).
The sums of all components are shown by solid curves (black).}
  \label{fig:fit}
 \end{center}
\end{figure}

The results of the fits of the above PDFs to the final data samples are shown in Fig.~\protect\ref{fig:fit}.
The signal yields and the reconstruction efficiencies are listed in Table~\protect\ref{tb:summary}.
Note that the rare charmless $b\rightarrow s$ decay $B^- \rightarrow K^+K^-\pi^-$
can peak inside the signal region for $B^-\rightarrow [K^+\pi^-]_DK^-$ and be included in the signal yield.
To estimate its contribution as well as contributions
from $B^-\rightarrow [K^+K^-]_D\pi^-$ and $B^-\rightarrow [K^-\pi^+]_DK^-$,
we fit the $M(K\pi)$ data sidebands:
$1.815~{\rm GeV}/c^2 < M(K\pi) < 1.845~{\rm GeV}/c^2$
and $1.885~{\rm GeV}/c^2 < M(K\pi) < 2.005~{\rm GeV}/c^2$.
The sidebands are chosen to avoid the contribution from $B^-\rightarrow [K^+K^-]_D\pi^-$
caused by $K/\pi$ misidentification.
We apply the same fitting method used in the signal extraction to the sideband sample
to obtain an expected yield of $-1.9^{+3.7}_{-3.5}$ events.
For $B^-\rightarrow [K^+\pi^-]_D\pi^-$, we also apply the requirement $M(K\pi) < 1.915~{\rm GeV}/c^2$
for the sideband sample to avoid $B^-\rightarrow [\pi^+\pi^-]_D\pi^-$ background, and obtain $-3.2^{+7.0}_{-6.4}$.
We do not subtract these backgrounds from the signal yields
but instead include the errors on the yields in the systematic uncertainties.

\begin{table}[htb]
 \caption{Signal yields, reconstruction efficiencies and significances.
 Charge-conjugate modes are included.
 The uncertainties shown are statistical only.}
 \label{tb:summary}
 \begin{center}
  \begin{tabular}{cccc}
  \hline \hline
  Mode & Yield & Efficiency (\%) & Significance \\ \hline
  $B^- \rightarrow [K^+\pi^-]_D K^-$ & $56.0^{+15.1}_{-14.2}$ & $33.6\pm0.4$ & $4.1\sigma$ \\
  $B^- \rightarrow [K^-\pi^+]_D K^-$ & $3394^{+68}_{-69}$ & $33.2\pm0.4$ & \\
  $B^- \rightarrow [K^+\pi^-]_D \pi^-$ & $165.0^{+19.1}_{-18.1}$ & $36.5\pm0.4$ & $9.2\sigma$ \\
  $B^- \rightarrow [K^-\pi^+]_D \pi^-$ & $49164^{+245}_{-244}$ & $35.7\pm0.4$ & \\
  \hline \hline
  \end{tabular}
 \end{center}
\end{table}

From the signal yields in Table~\ref{tb:summary}, we obtain
\begin{eqnarray}
  {\cal R}_{DK} &=& [1.63^{+0.44}_{-0.41}({\rm stat})^{+0.07}_{-0.13}({\rm syst})] \times10^{-2}, \\
  {\cal R}_{D\pi} &=& [3.28^{+0.38}_{-0.36}({\rm stat})^{+0.12}_{-0.18}({\rm syst})] \times10^{-3},
\end{eqnarray}
where the contributions to the systematic uncertainties are listed in Table~\protect\ref{tb:syst}.
The uncertainties due to the $\Delta E$ PDFs for the $DK$ signal, the $D\pi$ signal,
and the $D\pi$ feed-across
are evaluated by varying the shape parameters by $\pm 1 \sigma$.
Those due to the $DK$ feed-across are obtained by varying the width and the mean by $\pm 10\%$,
which is the difference observed between the data and MC samples for the $D\pi$ feed-across.
The uncertainties from the $\it NB$ PDFs
for the $DK$ and $D\pi$ signals (the $D\pi$ feed-across) are
estimated by obtaining PDFs from the region $0.01~{\rm GeV}<|\Delta E|<0.02~{\rm GeV}$
($0.01~{\rm GeV}<|\Delta E-0.05~{\rm GeV}|<0.02~{\rm GeV}$).
Those due to the $DK$ feed-across and the $B\bar{B}$ background
are estimated by using the $DK$ signal PDF. 
Those due to the $q\bar{q}$ background are estimated
by using ($M_{\rm bc}$, $\Delta E$) different sidebands.
The uncertainties due to the $K/\pi$ misidentification probabilities for the feed-across backgrounds
are obtained by varying their values by their $\pm 1 \sigma$ errors.
The uncertainty due to the charge asymmetry of the $q\bar{q}$ background
is obtained by varying it by $\pm 0.02$ ($\pm 0.005$) for $DK$ ($D\pi$),
which is the uncertainty in the favored $DK$ ($D\pi$) signal.
A possible fit bias is checked by generating 10,000 pseudo-experiments.
The uncertainties in detection efficiencies mainly arise from MC statistics
and the uncertainties in the particle identification efficiencies.
The total systematic uncertainty is calculated by summing the above uncertainties in quadrature.

\begin{table}[htb]
 \caption{Summary of the systematic uncertainties.
 We use the notation ``$\cdot\cdot\cdot$" to denote negligible contributions.}
 \label{tb:syst}
 \begin{center}
  \begin{tabular}{lcccc}
  \hline \hline
  Source & ${\cal R}_{DK}$ & ${\cal R}_{D\pi}$ & ${\cal A}_{DK}$ & ${\cal A}_{D\pi}$ \\ \hline
  PDFs of $\Delta E$ & $^{+2.1}_{-1.8}\%$ & $^{+1.3}_{-1.2}\%$ & $\pm0.01$ & $\pm0.00$ \\ 
  PDFs of $\it NB$ & $^{+3.4}_{-3.0}\%$ & $\pm3.1\%$ & $^{+0.02}_{-0.01}$ & $\pm0.01$ \\ 
  $K/\pi$ misidentification & $\pm0.2\%$ & $\pm0.0\%$ & $\pm0.00$ & $\pm0.00$ \\ 
  Asymmetry of $q\bar{q}$ background & $^{+0.8}_{-0.9}\%$ & $\pm0.1\%$ & $\pm0.01$ & $\pm0.00$ \\ 
  Fit bias & $-1.1\%$ & $-0.5\%$ & $-0.01$ & $-0.00$ \\ 
  Peaking backgrounds & $-6.6$\% & $-4.2$\% & $+0.03$ & $+0.00$ \\
  Efficiency & $\pm 1.7$\% & $\pm 1.5$\% & $\cdot\cdot\cdot$ & $\cdot\cdot\cdot$ \\
  Detector asymmetry & $\cdot\cdot\cdot$ & $\cdot\cdot\cdot$ & $\pm 0.02$ & $\pm0.00$ \\ \hline 
  Total & $^{+4.4}_{-7.8}\%$ & $^{+3.7}_{-5.6}\%$ & $^{+0.04}_{-0.03}$ & $^{+0.02}_{-0.01}$ \\
  \hline \hline
  \end{tabular}
 \end{center}
\end{table}

The significances of ${\cal R}_{DK}$ and ${\cal R}_{D\pi}$
are estimated using the fit likelihoods by convolving asymmetric Gaussians
denoting the systematic uncertainties~\protect\cite{ADS_Belle},
and listed in Table~\protect\ref{tb:summary}.
The significance for ${\cal R}_{DK}$ is 4.1$\sigma$,
which constitutes the first evidence for the suppressed $DK$ decay.

\begin{figure}[htb]
 \begin{center}
  \includegraphics[width=8.4cm,height=8.4cm]{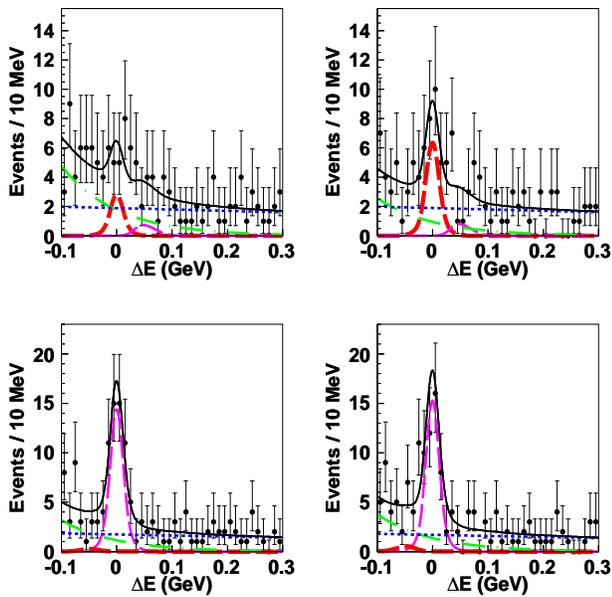}
  \caption{$\Delta E$ distributions (${\it NB}>0.9$) for $[K^+\pi^-]_DK^-$ (left upper), 
$[K^-\pi^+]_DK^+$ (right upper), $[K^+\pi^-]_D\pi^-$ (left lower),
and $[K^-\pi^+]_D\pi^+$ (right lower).
The curves show the same components as in Fig.~\protect\ref{fig:fit}.}
  \label{fig:fit_asym}
 \end{center}
\end{figure}

The $\Delta E$ projections are shown separately for each charge
of the $B$ candidate in Fig.~\protect\ref{fig:fit_asym}.
We obtain
\begin{eqnarray}
  {\cal A}_{DK} &=& -0.39^{+0.26}_{-0.28}({\rm stat})^{+0.04}_{-0.03}({\rm syst}),\\
  {\cal A}_{D\pi} &=& -0.04\pm 0.11({\rm stat})^{+0.02}_{-0.01}({\rm syst}),
\end{eqnarray}
where the systematic uncertainties are evaluated in a similar manner
as that done for ${\cal R}_{DK}$ and ${\cal R}_{D\pi}$ (see Table~\protect\ref{tb:syst}).
The uncertainty due to the yield of the peaking backgrounds
is obtained by varying the signal yield in the denominator of the $CP$ asymmetry.
The uncertainty due to the asymmetry of the peaking backgrounds is negligible~\cite{KKpi_BaBar}.
To account for possible bias due to the charge asymmetry of the detector,
we take the uncertainty in the asymmetry of the favored signal
as a conservative limit on this effect.

Assuming Eqs.~(\ref{eq:rdk_rel})--(\ref{eq:adk_rel}) and $r_B = 0.1$~\protect\cite{PDG},
the values of ${\cal R}_{DK}$ and $A_{DK}$ are restricted
to the ranges $[0.2,$ $2.5]\times 10^{-2}$ and $[-0.9,$ $0.9]$, respectively,
depending on the values of $\phi_3$, $\delta_B$, and $\delta_D$.
Our results are consistent with these expectations.
The small experimental uncertainties in ${\cal R}_{DK}$ and $A_{DK}$
thus provide important additional information on $\phi_3$.
The experimental results for ${\cal R}_{D\pi}$ and $A_{D\pi}$
are also consistent with the standard model~\protect\cite{PDG}.

In summary, we report measurements of the suppressed decay $B^- \rightarrow [K^+\pi^-]_Dh^-$ ($h = K, \pi$)
using the full $772\times 10^{6}$ $B\bar{B}$ pair data sample collected with the Belle detector.
We use a neural network-based method~\protect\cite{NeuroBayes}
to discriminate $q\bar{q}$ background from signal,
impose a $D^{*\pm}$ veto, and employ a two-dimensional fit to extract the signal.
These steps, along with a $20\%$ increase in the data sample,
result in a significant improvement
compared to the previous analysis~\protect\cite{ADS_Belle}.
We obtain the first evidence for a $DK$ signal with a significance of $4.1\sigma$,
and report the most precise measurements to date
of the $CP$ asymmetries and ratios of the suppressed decay rate to the favored decay rate.
Our results will provide important ingredients
in a model-independent extraction of $\phi_3$~\protect\cite{GLW, ADS}.

We thank the KEKB group for excellent operation of the
accelerator, the KEK cryogenics group for efficient solenoid
operations, and the KEK computer group and
the NII for valuable computing and SINET3 network support.  
We acknowledge support from MEXT, JSPS and Nagoya's TLPRC (Japan);
ARC and DIISR (Australia); NSFC (China); MSMT (Czechia);
DST (India); MEST, NRF, NSDC of KISTI, and WCU (Korea); MNiSW (Poland); 
MES and RFAAE (Russia); ARRS (Slovenia); SNSF (Switzerland); 
NSC and MOE (Taiwan); and DOE (USA).


\begin{thebibliography}{99}
\bibitem{gamma}
This angle is also known as $\gamma$.
\end{thebibliography}
\end{document}